\documentstyle[aps,prb,preprint,epsfig,floats]{revtex}

\begin{document}
\title{Notes on RVB-Vanilla by Anderson et al.}
\author{C.M.Varma}
\address{University of California, Riverside, Ca. 92521}
\maketitle
\begin{abstract}
The claims made for the Resonating Valence Bond ideas for the Cuprates 
in a recent paper by Anderson et al. on the basis of a variational calculation
are discussed.
\end{abstract}

\section{Introduction}

Within weeks of the confirmation of the discovery of high temperature superconductivity
 in the cuprates by Bednorz and Muller, Anderson\cite{pwa1} suggested an explanation of 
the phenomena and called it resonating valence bonds (RVB). Despite an enormous 
theoretical effort by the international scientific community, systematic or 
 consistent theoretical results have been hard to obtain on this 
idea for the model proposed by 
Anderson for the cuprates. When some
 specific predictions were made based on the general ideas or approximate calculations, 
 experiments did not conform. 

 In a curious recent note
 Anderson, Lee, Randeria, Rice, Trivedi and Zhang (ALRRTZ) \cite{pwa2}.
 put forth
 that the variational calculations that were done over a decade ago based on 
Anderson's ideas  and have been revived \cite{randeria} recently, support the 
ideas of RVB for the cuprates. They also reiterate that the Hubbard
 /t-J model, also proposed by Anderson is a sufficient model for the essential
 physics of the cuprates.

In this note I point out the following:

(1) The principal result  of the variational calculations discussed by ALRRTZ
 is contrary to a vast array of  experimental results. This disagreement is of a 
fatal nature for ideas which the variational calculation is taken to support.

(2)  The properties of the Hubbard/t-J model cannot be adequately studied
 using the wavefunctions with the limited
variational degree of freedom employed by ALRRTZ.

(3) In view of (1) and several other experimental and theoretical results,
 the Hubbard/t-J model is itself
inadequate as a model for  the Cuprates.

\section{Experiments and the results from the chosen 
 Wavefunction}

The chosen variational wavefunction  is the d-wave superconducting wavefunction with 
projection to remove double occupation:

\begin{eqnarray}
\Phi= \bar{P}\phi_{BCS}\left( \Delta({\bf k})\right).
\end{eqnarray}

$\bar{P}$ is the projection operator and $\Delta({\bf k}) $ is the the d-wave BCS 
order parameter function with a variational parameter of magnitude $\Delta_0$.
The principal result of the calculations is that the ground-state energy is
minimized when $\Delta_0$ varies with
$x$, the deviation from half-filling as shown schematically in figure (1), while the
 superconducting 
long-range order parameter $|\Psi|$ has a dependence with $x$ as also sketched 
in the figure. $\Delta_0$ is interpreted to represent the magnitude of the  
experimentally observed
pseudogap. A corollary to fig. (1) is a phase diagram sketched in fig. (2),
where $T_p^A$ marks the crossover temperature to pseudogap properties. 
The validity of the argument can be tested by comparing the results  
of figs. (1) and (2) with experiment. Other comparisons with experiment 
are not meaningful
if this test fails. 

\begin{figure}
\begin{center}
\includegraphics[width=8.3cm]{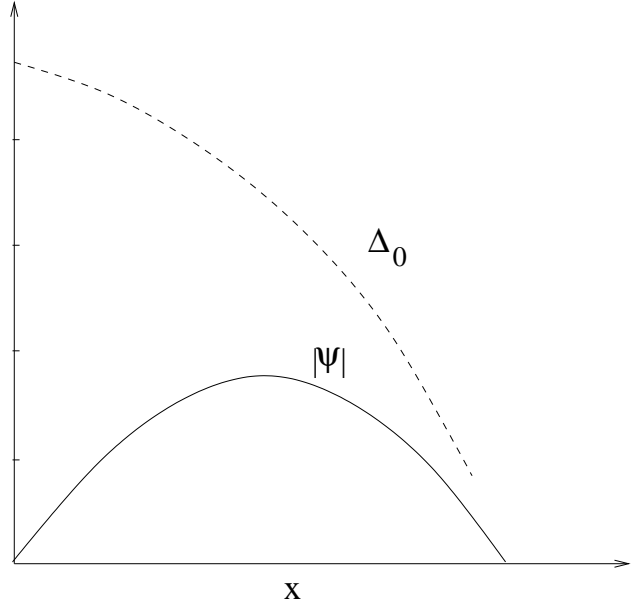}
\end{center}
\caption{The calculated variation  of the parameter $\Delta_0$ in the wavefunction
Eq. (1) and the calculated magnitude of the superconducting order parameter with x; adapted from Ref. (2)} 
\label{fig:tc}
\end{figure}

\begin{figure}
\begin{center}\includegraphics[width=8.3cm]{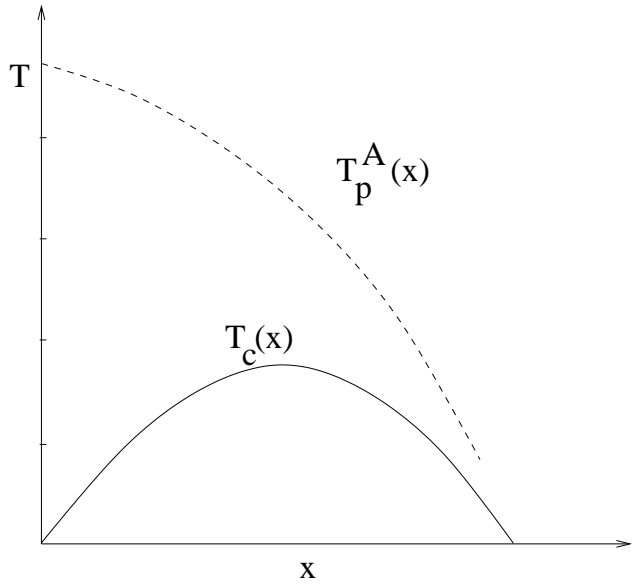}
\end{center}
\caption{The Phase diagram in the $ T-x$ plane implied by calculations reported in
Ref. (2)} 
\label{fig:delta}
\end{figure}

The authors acknowledge that the wavefunction does not correctly give the region near
 $x=0$, where the model studied as well as the Cuprates have an AFM ground state.  An
 unasked question is : At what x does the chosen RVB variational
 wavefunction have a lower energy than a wavefunction describing AFM at appropriate
 wave-vectors ${\bf Q}(x)$? 
They also acknowledge that their approach is no help in understanding the
 universal normal state properties for compositions near those for the highest $T_c$.
 I will return to both these points; first,
let us compare the claims made with the experiments.

 A crucial feature of  Fig. (1) is that the parameter $\Delta_0$ is finite throughout 
the superconducting region $x_{min}<x<x_{max}$. I summarize evidence below that,
in experiments, the pseudogap properties 
 disappear above a critical composition $x_c$ within the superconducting region.
 Moreover experiments show that a singularity exists at $x_c$, the Quantum Critical Point
(QCP) in the 
limit $T \rightarrow 0$. If $T_c$ is reduced by application of a magnetic field,
 the normal state anomalies continue to lower temperatures. The pseudogap region as
well as the Fermi-liquid region emanate from $x_c$. 
Any theory which is smooth across $x=x_c$ can then not be a theory for the 
Cuprates. The universal phase diagram of the cuprates \cite{cmv1}
 is as sketched in fig. (3).

The evidence is diverse and mutually consistent:

(1) Tunneling Measurements: The most direct is the measurement of pseduogap by tunneling
by Alff et al. \cite{alff} in two electron doped superconductors by reducing $T_c$
 on applying a magnetic field. The central observations are 
(a) for $x<x_c$ within the superconducting dome, the characteristic pseudogap feature
 in tunneling conductance is observed while
the superconducting gap feature disappears as $T_c(H)$ is reduced to 0, (b) This
feature appears at a temperature $T_p(x)$ below $T_c(x)$ at $H=0$,
(c) the magnitude of 
the pseudogap for three different samples
with $T_p(x)$ below $T_c$, and $T_p(x)$ itself, extrapolate to zero at an $x_c$ inside
 the superconducting dome. 

(2) Resistivity near $T=0$: In the region above $T_c(x)$ and below a 
temperature characterized by $T_p(x)$, a change in the temperature dependence
of the resistivity from linear to a higher power is observed. Similarly below a 
temperature $T_F(x)$ for $x>x_c$, a change in the power law tending towards the Fermi-
liquid value of 2 is observed. This is observed in all the cuprates studied. 
Fig.(4) organizes the data in a number of compounds \cite{panago}.

A linear temperature dependence of resistivity at low temperatures cannot occur
without a putative singularity in the fluctuation spectra at $T=0$. Measurements in
a large enough magnetic field to drive $T_c$ to 0 show that the linearity persists in an 
increasingley narrower region of $x$ as $T_c$ is decreased and persists at least down to
$40 mK$ in one cuprate compound \cite{greene} and down to at least 2 K in another 
\cite{boebinger}. To within the finest composition variation studied, $\delta x=0.01$, 
the resistivity power law changes to higher values on either side of $x_c$.

\begin{figure}
\begin{center}
\includegraphics[width=8.3cm]{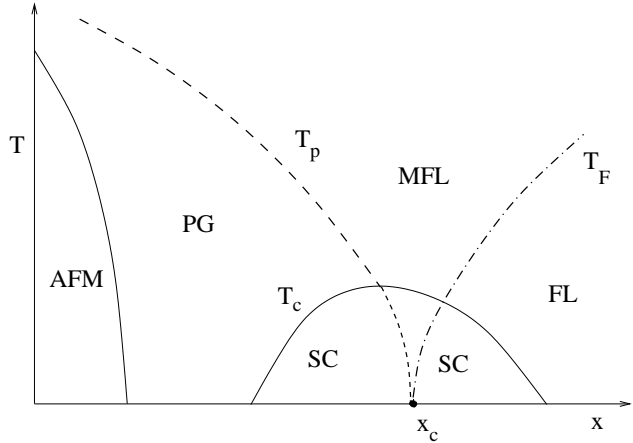}
\end{center}
\caption{The {\it universal} Phase diagram for the Cuprates; from Ref.(4)} 
\label{fig:phase}
\end{figure}

(3) Hall effect near $T=0$ \cite{greene}:  In the case of one of the class of
  compounds whose $T_c$ is reduced to $0$ by a magnetic field, the Hall coefficient
 has been measured at low temperatures. It shows a singular derivative
at $x \approx x_c$.

(4) Transport properties above $T_c(x)$: Actually, it is not necessary to study the
 properties near $T=0$ to rule out a pseudogap region beyond
an $x_c$ in the superconducting range of $x$. If fig. (2) were true, the universal 
normal state 
anomalies would change to the pseudogap properties for any $x$ for temperatures 
below $T^A_p(x)$ and above $T_c(x)$.  The data does not sustain this point
 of view.  The resistivity data consistent 
with the phase diagram (3) has  been shown in fig. (4). Wuydt et al.\cite{wuydt} 
have shown that the determination of $T_p(x)$ by thermodynamic measurements, 
specific heat and magnetic susceptibility
as well as the Cu-nuclear relaxation rate is consistent with that from resistivity. 
Recent measurements of the optical conductivity \cite{timusk} for various compositions
 are also 
consistent with the phase diagram (3).

In a very recent paper, Naqib et al.\cite{tallon} have measured the resistivity of 
$Y_{1-x}Ca_xBa_2(Cu{1-y}Zn_y)_3O_{7-\delta}$ over a wide range of $x$ and $y$ and in 
a magnetic field. Just as a magnetic field, $Zn$ reduces $T_c$ without
 affecting the pseudogap. They can thereby observe the variation of $T_p$
below the un-Zn-doped and zero magnetic field $T_c$. $T_p$ continues below 
this $T_c$, extrapolating to a finite value well within the superconducting dome.

\begin{figure}
\begin{center}
\includegraphics[width=8.3cm]{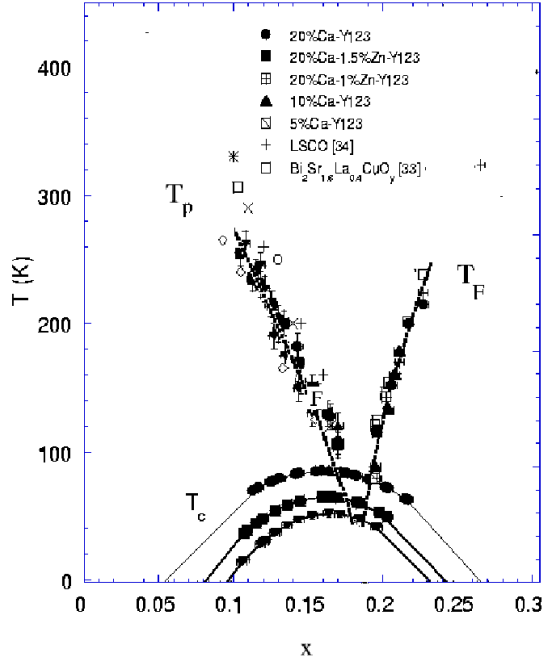}
\end{center}
\caption{Organization of the Resistivity data for a number of Cuprates: In the 
region between the set of points indicated by $T_c, T_p, T_F$, the resistivity is linear 
in temperature. It has a different dependence in other regions. Figure taken from  
Ref.(8).} 
\label{fig:panago}
\end{figure}

(5)Thermodynamic Measurements: ALRRTZ \cite{pwa2} quote only the thermodynamic 
measurements and analysis by Loram et al.\cite{loram}  both in the normal state and 
for the superconducting condensation energy, and claim not to understand how the 
conclusions were reached.
Actually the conclusions of Loram et al. are consistent with the existence of a QCP 
and all the experiments quoted above.This is discussed further below in 
connection with fig.(5).

(6) Raman Susceptibility: Careful measurements of Raman intensity in the 
compound 123 near $x=x_{opt}$, the composition for the highest $T_c$, 
show \cite{klein}  that the susceptibility
has the scale invariant form
\begin{eqnarray}
Im~\chi(\omega/T)~& \propto&~ \omega/T,~~ for ~\omega/T~ \ll~1  \\ \nonumber 
                                    &\propto&~~constant, ~~ for~ \omega/T~~\gg ~1.
 \end{eqnarray}
 More complete are measurements \cite{cardona} in the compound 248. At 
stoichiometry, this compound displays the pseudogap properties with a
 characteristic change in resistivity from linear to that of the pseudogap
 behavior below $T_p \approx  200 K$  and a $T_c$ of 80 K. Under pressure P,
 $T_c$ continuously rises to 110 K at  a P of 100 kbar; simultaneously $T_p$ 
decreases and is invisible above a P about 80 kbar \cite{rin248}. Raman measurements
 under pressure reveal the form of Eq. (2) near 100 kbar but at lower P one finds a
 characterstic infra-red cut-off proportional to $T_p(P)$. 

In their paper ALRRTZ reproduce  the magnitude of a gap deduced
 from the photoemission experiments \cite{campu} which indeed varies with
 $x$ in the manner similar to fig. (1). This gap is deduced at low temperatures 
from experiments in the superconducting phase. This may appear reasonable enough.
 After all the comparison is being made to 
a parameter in a ground state wavefunction. Actually, this is a specious 
argument if $\Delta_0$ is understood to represent the pseudogap.
A tunneling measurement made below $T_c$ will show a gap which is zero only when
both the pseudogap and the superconducting gap are zero because the gap measured is 
is an appropriate combination of the 
pseudogap $D(x)$ and the superconducting gap $\Delta_{sc}(x)$; for instance
\begin{eqnarray}
gap = \sqrt{|D({\bf k})|^2+|\Delta_{sc}({\bf k})|^2}
\end{eqnarray}
To measure the pseudogap one must kill the superconducting gap as Alff et al. \cite{alff}
have done, as discussed above. The total gap in zero field and at low temperatures
 then goes to zero only for 
$x>x_{max}$. Alternately one may deduce the pseudogap from measurements above $T_c$. 
Loram et al.
have done this through modelling their thermodyanmic measurements. 
A comparison of their deduced values of the pseudogap as a function of $x$ with those 
obtained from
 tunneling and photoemission at low
temperature is shown \cite{batlogg} in fig.(5). The two sets of gaps do follow a relation not
inconsistent with Eq. (3) if $\Delta_{sc0}$ is proportional to $T_c(x)$.

\begin{figure}
\begin{center}
\includegraphics[width=8.3cm]{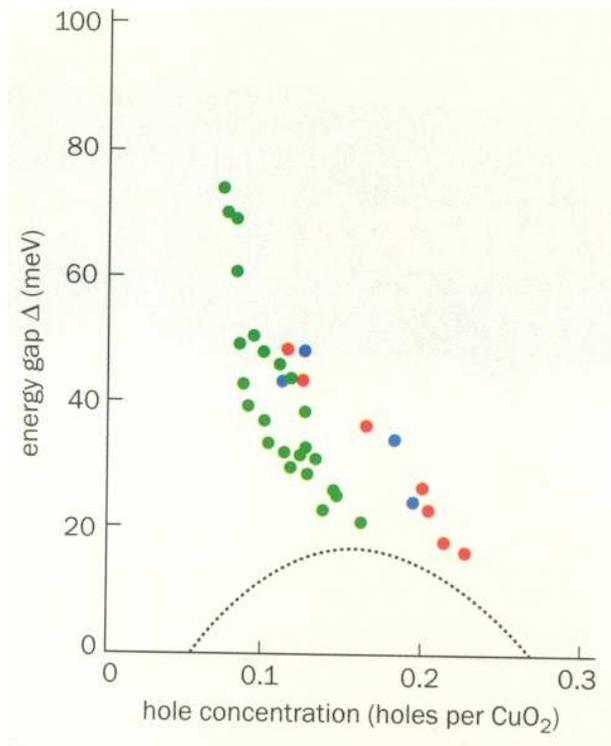}
\end{center}
\caption{The energy gap in the normal state determined by fitting to the specific heat
data with a model density of states by Loram et al. is shown in green. The tunneling
 experiments well below $T_c$ give the gap shown in red in agreement with ARPES
 shown in blue.}
\label{fig:pgap}
\end{figure}

\section{Restricted Variational freedom}
The variational results are at odds with the phase diagram for the cuprates. 
Are they a good representation of the physics of the t-J model? What does the 
variational parameter $\Delta_0$ really represent?
In a model with strong local interactions $U>>t$, there must be a transfer of 
the distribution function $n({\bf k})$ from below the Fermi-vector to above
 compared to non-interacting electrons. In the wavefunction, Eq. (1), 
$\Delta_0$ is the only parameter to accomplish this physics. There is 
then simply no choice but for the calculations to exhibit a pseudogap.
 It is put in by hand without comparing the ground state energy 
with other wavefunctions which accomplish the same physics. The choice of
Eq. (1)
automatically ensures
$\Delta_0 \approx J$ for $x\approx 0$ with a decline to $\approx 0$ for
 large enough $x$.  A Fermi-liquid,  accomplishes the same physics with
 a discontinuity $z$ in $n({\bf k})$ at ${\bf k}_F$. $z$ changing smoothly 
from $0$ near $x=0$ to $1$ for large enough $x$.  Much more relevant for the
 physics of the t-J model would be to compare the ground state energy using
 Eq. (1) with that for a wavefunction representing AFM order at a wave-vector
 ${\bf Q}(x)$, also projected to remove double-occupation. 
We know the wavefunction (1) is wrong for the t-J model near $x=0$ only because
 the experiments say the region near $x=0$ is AFM.  How far does this problem 
with the wavefunction persist. Only comparison of the energy with an AFM wavefunction
 can tell. Based on the results of the mean-field calculations and several numerical
 calculations,  my conjecture is that the regime of AFM for the t-J model is much more 
extended than in the experiments and that in variational calculations the phase diagram 
will show AFM followed by a co-existing region of AFM and d-wave superconductivity
 followed by d-wave superconductivity alone. The physics is much the same as that 
for d-wave superconductivity derived from AFM fluctuations
with numerical corrections from Fermi-liquid effects \cite{miyake}.

This is not to say that the t-J model has a superconducting ground state above 
some $x$. As Anderson et al. point out,  numerical evidence on this issue is 
divided.  
 But, from the perspective of a theory of the cuprate phenomena this issue is 
incidental. Even if the ground state of the t-J model were to be superconducting,
 it cannot be a model for the Cuprates unless it can give a phase diagram similar 
to Fig. (3). 

\section{ Model for Cuprates}

It was suggested \cite{cmv2} that a unique property of the cuprates is that the
 ionization energy of the $Cu^{++}$ ions is very close to the affinity energy of
 $O^{--}$. (This is what makes the half-filled Cuprates {\it charge transfer 
insulators}\cite{zaanen}). Their difference $\epsilon_{pd}$ is similar to  
 the kinetic energy parameter $t_{pd}$ and estimates of the screened nearest neighbor
 Cu-O ionic interactions V.  In this condition, in the metallic phase charge
 fluctuations exist almost equally on Cu and O ions and that new physics can 
arise in a model with three dynamic degrees of freedom per unit cell (two Oxygens 
and one Copper)due to the ionic interactions. Anderson et al. argue instead that as observed in experiments
 there is only one band near the Fermi-surface and in any case the three orbital
 model may be reduced to a one effective orbital model by a canonical transformation. 
In this canonical transformation, they ignore the interactions represented by V. 
It is easy to show that for $zV\geq t_{pd}, \epsilon_p-\epsilon_d$, where z is the number of 
nearest neighbors, this canonical
transformation does not converge.

The argument has never been that there is more than one band near the chemical
 potential. The relevant question is what is the nature of the wavefunctions 
in this band and the effective interactions among one-particle excitations of
 the band after the states far from the chemical potential (in band calculations)
 are eliminated. That new physics arises in the general model
(3-orbitals and ionic interactions
as well as local interactions) \cite{cmv2}
 is known from two exact solutions : (a) A local model, which
 bears the same relation to the general model as the Anderson model for local
 moments bears to the Hubbard model. A QCP is found in the model with logarithmic
 singularities unlike the Ground state singlet of the
Anderson or Kondo model \cite{perakis}. (b) Long one -dimensional
 ----Cu-O-Cu-O--- chains on which extensive numerical results 
\cite{sudbo} have been obtained. This model gives new
 physics including long-rage superconducting correlations  
not found when the ionic interactions are put to 0.

Mean-field calculations \cite{cmv2} on the general model give 
a phase diagram similar to fig. (3), with a QCP at $x_c$ and an 
unusual time-reversal violating phase for the pseudogap region.

\section{Concluding Remarks}

  If fluctuations of the form of Eq. (2) persist down to
$T\rightarrow 0$, they give that the real part of the susceptibility has the 
singularity
\begin{eqnarray}
Re \chi (\omega.T) ~&~\propto ~&~\ln (\omega_c/\omega)~~for~ T\ll \omega,\\ \nonumber
                                    &~ \propto~&~ \ln (\omega_c/T)~~for~ T\gg \omega,
\end{eqnarray}
assuming a high energy cut-off $\omega_c$.
I singled out the Raman  experiments in Sec. 2, because they  directly  measure the 
singularity  in the region near $x_{opt}$. They also show the crossover to fluctuations
 with an infrared cut-off in the pseudogap region. It is true that the measurements are
 at present available only above $T_c(x)$. But, given all the other experimental results,
 can there be any doubt that when measurements are carried out in a magnetic field
to reduce $T_c$ to $0$, the singularity of Eqs. (2,4) will continue at $x_c$ with an 
infrared cut-off on both sides of it. 

This singularity specifies not only  a QCP in the Cuprates inside the superconducting
 dome but also the critical exponents of the fluctuations about it. This form was 
predicted \cite{cmv3} in 1989 on phenomenological grounds. Actually what is measured
 in Raman scattering is only the long wavelength limit of the predicted form.  
Essentially every normal state anomaly in region I has the marginal Fermi-liquid form 
which  follows from this singularity.
The lineshape in single-particle spectra as a function of momentum and energy was 
predicted and verified in ARPES experiments. From  two parameters, obtainable from 
the ARPES spectra quantitative agreement with transport properties including the 
Hall effect have been obtained to within a factor of 2. 

But what about superconductivity? In any theory based on Cooper pairs, the 
fluctuations in the normal state and their coupling to the fermions engender 
the superconducting instability.  The high frequency cut-off and the coupling 
constant of the singular fluctuation obtainable from normal state properties
 certainly give the right order of magnitude of $T_c$. This idea can be 
quantitatively tested in complete detail through an analysis of ARPES 
experiments in the superconducting state \cite{vekhter}  which is an extension
 to anisotropic superconductors of the Rowell-Mcmillan method of analysing 
tunneling data for s-wave superconductors \cite{footnote1}.  

What of the microscopic theory for the singularity and the fluctuations about it? 
A QCP in more than one-dimension is generally the end-point of a line of phase 
transitions. From Raman scattering results, as well as lack of any convincing 
evidence for change of a translational or spin-rotational symmetry, we know that
 the change of symmetry at this phase transition must occur at $q=0$ and  in the 
spin-singlet channel. Based on the general model for Cuprates, an unusual
 Time-reversal violating phase has been predicted for the pseudogap phase
 \cite{cmv1}. A new experimental technique using ARPES was suggested to look for
 this phase \cite{cmv4}.
This experiment \cite{kaminski} has been successful. A confirmation of this phase
 by this or other methods should remove any doubt on what is the minimum model for
 the Cuprates and what is the nature of the fluctuations responsible for the Cuprate
 phenomena \cite{z2}.

Arguments have been given why the fluctuations to this phase in this model produce 
the specified fluctuations. An exact calcualtion of the related local model \cite{perakis}
produces precisely the specified form of fluctuations. More work on this issue will be
forthcoming.

The vast array of experiments in the cuprates narrows the possible theories for the 
phenomena.  One can be certain that the theories should have a QCP with a phase with
a pseudogap ending at it. One can also be 
certain that in the long wavelength limit, the fluctuations about the QCP 
must have the form of
 Eq. (2) because that is what is observed \cite{q-dep}. 

Unless the ideas of RVB satisfy these requirements, they together with several others 
which do not, can be excluded as a framework for a theory of the cuprate phenomena. 
 The first paper 
of Anderson was important
for cuprates;  it posited the phenomena was due to  electron-electron interactions and 
suggested that quite new physics will be required to understand it. This has turned out 
to be true.
Its  direct importance is to physics other than that in cuprates, in the interest it has 
provoked in search for models in which quantum-mechanics leads to beautiful new 
states of existence such as RVB.
This may have implications for physics yet to come.


\begin{references}
\bibitem{footnote1}
Since $T_c$ is determined by the spectra at $T_c$,
the analysis of single-particle self-energy just above $T_c$ near $x_{opt}$ does not 
leave much  doubt that  the analysis just below $T_c$ will yield that the glue for 
superconductivity has the fluctuation spectrum of Eq. (2) at essentially all $q$. 
Such experiments and analysis should nevertheless be done to be certain.
\bibitem{z2}
The principal reason (and a valid one) for scepticism in thinking of the 
Pseudogap region as one of 
some broken symmetry is that, although most thermodynamic and transport 
properties change at $T_p(x)$, there is no  observable singularity in 
the specific heat at $T_p(x)$. Such a situation is  not excluded by
 statistical mechanics. The symmetry change observed 
in Ref.(\onlinecite{kaminski}) belongs to a class of $Z2\otimes Z2$ models,
which in appropriate range of parameters do have a phase transition 
without a diverging specific heat. This model should however give at least a bump
in the specific heat. Since, even that is not observed,
I suspect there are some more subtle issues involved.
\bibitem{q-dep}
It is also fairly certain
that the momentum dependence of the fluctuations must be much less 
important than their frequency dependence because that is the only way known to have
 the transport scattering rates be similar to the single-particle relaxation rates
and to understand the NMR experiments on cu in the normal state.A specific form with
dyanmical critical exponent of infinity has been proposed.
\end{references}
 \end{document}